\documentclass[11pt]{article}
\usepackage[T1]{fontenc}
\usepackage{lmodern}
\usepackage[a4paper,margin=25.4mm]{geometry}
\usepackage{amsmath,amssymb,amsthm,mathtools,microtype}
\usepackage{xcolor}
\usepackage[colorlinks=true,linkcolor=blue!45!black,citecolor=blue!45!black,urlcolor=blue!45!black]{hyperref}
\hypersetup{pdftitle={One Strategic Woman: Sharp Ranks and a Persistent Core},pdfauthor={Ron Kupfer}}
\newtheorem{theorem}{Theorem}
\newtheorem{lemma}{Lemma}
\newtheorem{corollary}[theorem]{Corollary}
\newcommand{\E}{\mathbb E}
\newcommand{\PP}{\mathbb P}
\newcommand{\one}{\mathbf 1}
\newcommand{\SM}{\mathcal S}
\title{One Strategic Woman:\\Sharp Ranks and a Persistent Core}
\author{Ron Kupfer}
\date{}
\begin{document}
\maketitle
\begin{abstract}
In a balanced random stable-marriage market, one fixed woman manipulates men-proposing deferred acceptance to obtain her best attainable partner. We prove that the women's expected average true partner rank is asymptotic to $(\ln n)^2$, improving the previous $O((\ln n)^4)$ bound, while the men's is asymptotic to $n/(2\ln n)$. The men's and women's average ranks, normalized by $n/\ln n$ and $\ln n$, converge jointly in distribution to $(U,U^{-1})$, where $U$ is uniform on $(0,1)$. Under optimal truthful truncation, the span of women's average ranks within the reported stable core, divided by $\ln n$, converges in distribution to $U^{-1}-1$.
\end{abstract}

\section{Results and proof outline}

In a balanced market with independent uniform preferences, men-proposing deferred acceptance (DA) gives the men average partner rank about $\ln n$ and the women about $n/\ln n$~\cite{pittel89}. Kupfer~\cite[Theorem~1]{kupfer} showed that one optimally strategic woman changes the women's expected average rank to $O((\ln n)^4)$. We determine both sides' sharp mean ranks and joint limit, and quantify the stable core that remains.

There are $n$ men and $n$ women with independent uniform strict complete preferences; every partner is truly acceptable. A woman $g$, fixed before preferences are drawn, observes the profile and submits an arbitrary ordered list, declaring omitted men unacceptable. All others are truthful. A report is \emph{optimal for $g$} if it gives her the best attainable partner under DA. Rank~1 is best. Let $R^{\mathrm M}$ and $R^{\mathrm W}$ be the men's and women's average true partner ranks, with their dependence on $n$ implicit. All logarithms are natural.

\begin{theorem}\label{thm:ranks}
Under any report optimal for $g$,
\begin{equation}\label{eq:means}
\E R^{\mathrm W}\sim (\ln n)^2,\qquad \E R^{\mathrm M}\sim\frac{n}{2\ln n}.
\end{equation}
If $U$ is uniform on $(0,1)$, then
\begin{equation}\label{eq:joint}
\left(\frac{\ln n}{n}R^{\mathrm M},\frac{R^{\mathrm W}}{\ln n}\right)
\xrightarrow{d}(U,U^{-1}).
\end{equation}
\end{theorem}

The average $R^{\mathrm W}$ is taken over the women in one market; its expectation also averages over random preference profiles. The women's limit in~\eqref{eq:joint} is Pareto: $\PP(U^{-1}>x)=1/x$ for $x\ge1$. Thus the median of $R^{\mathrm W}$ is asymptotic to $2\ln n$, while rare poor outcomes raise its expectation to $(\ln n)^2$. At the women-optimal matching, the expected women's average rank is $\sim\ln n$~\cite{pittel89}; Section~\ref{sec:core} quantifies the comparison. The coefficient $1$ corrects the heuristic $1/2$ in~\cite[Section~3.2]{kupfer}.

The proof uses an auxiliary DA run in which $g$ rejects every proposal, revealing $K$ distinct proposers. The optimal truthful truncation follows this run until her favorite among them proposes, then accepts him permanently. His index $J$ is conditionally uniform on $[K]$. Independent insertion shows that $K\sim n/\ln n$ and offer $j$, beyond a short initial phase, arrives after about $nj$ proposals. Pittel's rank hyperbola then gives $R^{\mathrm M}\sim J$ and $R^{\mathrm W}\sim n/J$. Uniform and harmonic averaging yield~\eqref{eq:means}. Section~\ref{sec:core} gives the core consequences.

We write $H_k=\sum_{i=1}^k1/i$. A failure probability $n^{-\omega(1)}$ is smaller than every fixed inverse power of $n$. Uniform $o(1)$ errors on good events are bounded by deterministic sequences tending to zero.

\section{A uniform index for the optimal outcome}\label{sec:representation}

We use the rejection-process viewpoint of~\cite[Sections~2 and~3.3]{kupfer}. Call women other than $g$ \emph{ordinary}. Men propose in their full true order, counting proposals immediately rejected as unacceptable. In the \emph{auxiliary empty-list run}, $g$ rejects everything; select the smallest-labeled free man with an untried woman. The run ends with every ordinary woman matched and one man unmatched and exhausted. Write $\mu_0$ for this reduced stable matching and $S=(b_1,\ldots,b_K)$ for $g$'s proposers in order. Let $b^*$ be her true favorite in $S$, at index $J$.

\begin{lemma}\label{lem:representation}
Truncating $g$'s true list below $b^*$ gives a perfect true-stable matching $\mu^*$. The man $b^*$ is her best attainable and best stable partner, and every optimal report yields $\mu^*$.
\end{lemma}
\begin{proof}
Under any report by $g$, each man's proposals form a prefix of his empty-list proposals. Otherwise consider the first rejection by a man's $\mu_0$ partner. She holds a preferred man $c$, who has not passed his own $\mu_0$ partner, or is unmatched in $\mu_0$. Either case blocks $\mu_0$. The unmatched man's empty-list proposals include everyone, so he is no exception.

The truncation follows the empty-list run until $b^*$ proposes, accepts him, and never loses him: every possible proposer belongs to $S$. Its outcome is perfect, since an unmatched man would eventually propose to any unmatched ordinary woman. It is true-stable: every rejected proposer to $g$ is worse than $b^*$, and men who never proposed prefer their final wives. Containment also proves optimality. If $c$ is any true stable partner, truncation at $c$ preserves a perfect stable matching, so DA gives $g$ an acceptable man in $S$. Hence $b^*$ is her best stable partner.

Any optimal-report outcome $\nu$ assigns $b^*$ to $g$. The same unmatched-agent argument makes $\nu$ perfect, and containment makes it true-stable. Thus $\nu$ is stable under the truncation, so every man weakly prefers $\mu^*$ to $\nu$. Any blocking pair of $\mu^*$ under the arbitrary report would involve $g$ and would also block $\nu$. The reverse men-optimality comparison gives $\nu=\mu^*$.
\end{proof}

These arguments use standard DA properties~\cite{gs62}; the equivalence of optimal reports is also a singleton case of~\cite[Theorem~6 and subsequent discussion]{shen18}. Conditional on all preferences except $g$'s true ranking, $S$ is fixed and
\begin{equation}\label{eq:index}
\PP(J=j\mid S,\text{empty-list run})=\frac1K\quad(j\in[K]).
\end{equation}
No stopping-time assumption about the informed optimum is needed.

Delete $g$ from the men's lists. Conditional on the reduced profile, recover their full lists by independent uniform insertion positions $I_m\in[n]$. A rejection by $g$ leaves the same man active, so the ordinary trajectory is unchanged. After $d$ ordinary proposals let $q_m(d)$ count man $m$'s proposals and set
\begin{equation}\label{eq:insert}
Z_d=\sum_m\one\{I_m\le q_m(d)\},\qquad
\E[Z_d\mid\text{reduced profile}]=d/n.
\end{equation}
The summands are conditionally independent. At the cut after ordinary proposal $d$, before any further proposal, exactly $Z_d$ offers have gone to $g$. If $d_j$ ordinary proposals precede her $j$th offer, then
\begin{equation}\label{eq:boundary}
j=Z_{d_j}+1.
\end{equation}
This includes a final offer by the exhausted man. Extend the exposure order after reduced termination in a fixed order preserving each man's list order, so that $Z_d$ is defined for all $d\le n(n-1)$.

\begin{lemma}\label{lem:counts}
Except with probability $n^{-\omega(1)}$, $K=(1+o(1))n/\ln n$ and, uniformly for $a=\lceil (\ln n)^{3/2}\rceil\le j\le K$, all ordinary women hold partners and $d_j+j=(1+o(1))nj$. Consequently,
\begin{equation}\label{eq:index-moments}
\E J\sim\frac{n}{2\ln n},\qquad
\E[n/K]\sim \ln n,\qquad \E[n/J]\sim (\ln n)^2.
\end{equation}
\end{lemma}
\begin{proof}
For $\varepsilon=(\ln n)^{-1/8}$ and $d_0=\lceil na/2\rceil$, conditional Chernoff and a union over all prefixes give
\[
\PP\left(\exists d\ge d_0:\left|Z_d-d/n\right|>\varepsilon d/n\right)
\le2n^2e^{-\varepsilon^2d_0/(3n)}=n^{-\omega(1)}.
\]
Monotonicity at $d_0$ rules out $j\ge a$ with $d_j<d_0$. Equation~\eqref{eq:boundary} now gives $d_j\sim nj$ uniformly.

For coverage, draw ordinary women uniformly and independently, discarding a man's repeated targets. A woman's first occurrence is nonredundant. Coupon collection gives coverage within $d_1=\lceil n(\ln n)^{5/4}\rceil$ draws, hence at most $d_1$ genuine proposals, with failure at most $ne^{-(\ln n)^{5/4}}$. Chernoff at $d_1$ gives $Z_{d_1}\le2(\ln n)^{5/4}<a-1$ with the same negligible failure. Thus indices $j\ge a$ occur after coverage.

Let $T$ be the terminal ordinary-proposal count. Pittel's unequal-market theorem~\cite[Theorem~1.5]{pittel19}, with sexes interchanged and side sizes $n-1,n$, gives $T\sim n^2/\ln n$ with failure $e^{-c(\ln n)^{3/2}}$. Indeed, its matched long-side rank center is $(n-1)^2(1/\ln n-1/(n-1))$; the unmatched man adds $n-1$ proposals. Relative error $O((\ln n)^{-1/4})$ is permitted.

At termination the unmatched man certainly proposes to $g$, while each matched man does so independently when $I_m\le q_m(T)$. Therefore
\[
K=1+\sum_{m:\,\mu_0(m)\ne\varnothing}\one\{I_m\le q_m(T)\},
\qquad \E[K\mid\text{reduced profile}]=\frac{T+1}{n}.
\]
Conditional Chernoff proves the claim about $K$. Now~\eqref{eq:index} gives $\E[J\mid K]=(K+1)/2$ and $\E[n/J\mid K]=nH_K/K$. Concentration of $K$, with $J,n/K,nH_K/K\le n$, proves~\eqref{eq:index-moments}.
\end{proof}

\section{Sharp mean ranks and the joint law}\label{sec:rank-proof}

Let $\SM$ be the set of true stable matchings. For $\mu\in\SM$, let $Q_{\mathrm M}(\mu),Q_{\mathrm W}(\mu)$ be the men's and women's total ranks. The classical inputs we need are, simultaneously for all such $\mu$,
\begin{equation}\label{eq:classical}
\left|Q_{\mathrm M}(\mu)Q_{\mathrm W}(\mu)/n^3-1\right|\le n^{-1/8},
\qquad Q_{\mathrm W}(\mu)\le2n^2/\ln n,
\end{equation}
with failure $n^{-\omega(1)}$. The hyperbola is due to Pittel~\cite{pittel92}, with its precise uniform statement recalled in~\cite[arXiv version, Section~1, p.~3]{pittel19}. The second bound follows by interchanging sexes in~\cite[Proposition, (4.4)]{pittel89}: for every $A$, the maximum total rank is at most $(n^2/\ln n)(1+(\ln\ln n+c_A)/\ln n)$ with failure $O_A(n^{-A})$.

\begin{proof}[Proof of Theorem~\ref{thm:ranks}]
For $J\ge a$ on Lemma~\ref{lem:counts}'s event, accepting $b_J$ matches the only free man and ends DA. Thus $Q_{\mathrm M}(\mu^*)=d_J+J\sim nJ$. True stability and~\eqref{eq:classical} yield, with uniform errors,
\begin{equation}\label{eq:coupling}
R^{\mathrm M}=(1+o(1))J,\qquad R^{\mathrm W}=(1+o(1))n/J.
\end{equation}
The hyperbola is applied only to the actual stable outcome; we never condition the uniform-index law on its good event.

Exceptional events contribute $o(1)$ to the expectations below, since $R^{\mathrm M},R^{\mathrm W},J,n/J\le n$. For the initial indices, \eqref{eq:index-moments} gives
\[
\PP(J<a)\le a\E[1/K]=O(a\ln n/n),\qquad
\E[n/J;\,J<a]\le H_a\E[n/K]=O(\ln n\ln\ln n).
\]
Using~\eqref{eq:classical} and $R^{\mathrm M}\le n$ gives
\[
\E[R^{\mathrm W};\,J<a]=O(a)=o((\ln n)^2),\qquad
\E[R^{\mathrm M};\,J<a]=O(a\ln n)=o(n/\ln n).
\]
Also $\E[J;J<a]\le a=o(n/\ln n)$. Averaging~\eqref{eq:coupling} and using~\eqref{eq:index-moments} proves~\eqref{eq:means}.

Finally $J/K\Rightarrow U$ by~\eqref{eq:index}, $K\ln n/n\to1$ in probability, and $\PP(J<a)=o(1)$. Substituting these relations in~\eqref{eq:coupling} proves~\eqref{eq:joint}.
\end{proof}

\section{The residual stable core}\label{sec:core}

Here $g$ uses the \emph{truthful truncation} at $b^*$; let $\mathcal C_g$ be the set of stable matchings for this reported profile. Other optimal reorderings have the same DA outcome but need not have the same reported core. Write $\mu^W$ for the original women-optimal matching. The standard constrained-core description~\cite[Section~3.1]{nnv21} gives, in the true stable-matching lattice ordered by the women's preferences,
\begin{equation}\label{eq:interval}
\mathcal C_g=\{\mu\in\SM:\mu(g)=b^*\}=[\mu^*,\mu^W].
\end{equation}
Define the extreme average ranks
\[
R_-^{\mathrm W}=\min_{\mu\in\mathcal C_g}Q_{\mathrm W}(\mu)/n,\qquad
R_+^{\mathrm W}=\max_{\mu\in\mathcal C_g}Q_{\mathrm W}(\mu)/n.
\]

\begin{corollary}\label{thm:core}
If $U$ is uniform on $(0,1)$, then
\begin{equation}\label{eq:core-ranks}
\left(\frac{R_-^{\mathrm W}}{\ln n},\frac{R_+^{\mathrm W}}{\ln n}\right)\Rightarrow(1,U^{-1}),
\qquad
\frac{R_+^{\mathrm W}-R_-^{\mathrm W}}{\ln n}\Rightarrow U^{-1}-1.
\end{equation}
In particular, $\PP(|\mathcal C_g|=1)\to0$.
\end{corollary}
\begin{proof}
By~\eqref{eq:interval}, $R_+^{\mathrm W}=R^{\mathrm W}$, whereas $R_-^{\mathrm W}/\ln n\to1$ in probability by the classical women-optimal rank estimate~\cite{pittel89}. Theorem~\ref{thm:ranks} and Slutsky's theorem give both limits. A singleton core has zero width, and the limiting width has no atom at zero.
\end{proof}

The manipulator receives her partner in $\mu^W$. For the women's average rank, $R^{\mathrm W}/R_-^{\mathrm W}\Rightarrow U^{-1}$; equivalently, $\PP(R^{\mathrm W}\le cR_-^{\mathrm W})\to1-1/c$ for each fixed $c\ge1$. Thus the ratio to the women-optimal average has a nondegenerate limit.

For cardinality, the lattice analysis of Ndiaye, Norin, and Vetta~\cite[Section~3.1 and Theorem~5.1]{nnv21} yields the comparison
\begin{equation}\label{eq:core-size}
n^{(\ln2)/2-o(1)}\le\E|\mathcal C_g|\le(1+o(1))e^{-1}n\ln n.
\end{equation}
Indeed, let $k$ count the stable matchings immediately above $\mu^W$ in the men's preference order. These matchings change disjoint nonempty sets of women. Joins of distinct subsets are distinct; omitting the at-most-one matching that changes $g$ gives $|\mathcal C_g|\ge2^{(k-1)_+}$. Since $\E k=\tfrac12\ln n+O(\ln\ln n)$, Jensen's inequality gives the lower bound. The upper bound uses $\mathcal C_g\subseteq\SM$ and Pittel's pre-manipulation estimate $\E|\SM|\sim e^{-1}n\ln n$~\cite[Theorem~1]{pittel89}. These bounds leave the asymptotic expected residual cardinality undetermined.

\end{document}